\documentclass[journal=acsnano,manuscript=article]{achemso}
\usepackage{graphicx}
\usepackage{multirow}
\usepackage{threeparttable}
\usepackage{dcolumn}
\usepackage{bm}
\usepackage{mathtools,amssymb}
\usepackage{upgreek}
\usepackage{color}
\usepackage{float}
\usepackage{pdfpages}
\usepackage[utf8]{inputenc}
\usepackage{mathptmx}
\usepackage{etoolbox}
\usepackage[symbol]{footmisc}
\usepackage{chemformula} 
\usepackage[T1]{fontenc} 
\usepackage{amsmath}

\makeatletter

\usepackage{amsfonts}
\usepackage{amsmath}
\usepackage{amssymb} 
\usepackage{bm} 
\usepackage{color}
\usepackage{enumerate}
\usepackage{graphicx}  
\usepackage{subcaption}  
\usepackage{caption}  
\usepackage[justification=justified,singlelinecheck=false]{caption}
\usepackage{dcolumn}
\usepackage{hyperref}
\usepackage{ragged2e}
\usepackage{lineno}

\hypersetup{
	colorlinks=true,       
	linkcolor=blue,          
	citecolor=magenta,        
	filecolor=cyan,      
	urlcolor=magenta           
}

\labelformat{section}{#1} 
\labelformat{subsection}{Section #1} 
\labelformat{figure}{Fig.~#1} 
\labelformat{subfigure}{Fig.~\thefigure#1}
\author{Vineet Pandey}
\affiliation[Indian Institute of Technology, Kharagpur]
{Materials Science Centre, Indian Institute of Technology, Kharagpur, West Bengal - 721302,India }

\author{Prasenjit Ghosh}
\affiliation[Indian Institute of Technology, Kharagpur]
{Materials Science Centre, Indian Institute of Technology, Kharagpur, West Bengal - 721302,India }

\author{Riju Pal}
\affiliation[S. N. Bose National Centre for Basic Sciences, Kolkata]
{S. N. Bose National Centre for Basic Sciences, Kolkata, West Bengal -700106,India }

\author{Sourav Paul}
\affiliation[Indian Institute of Technology, Kharagpur]
{Materials Science Centre, Indian Institute of Technology, Kharagpur, West Bengal - 721302,India }

\author{Abhijith M B}
\affiliation[Indian Institute of Technology, Kharagpur]
{Materials Science Centre, Indian Institute of Technology, Kharagpur, West Bengal - 721302,India }

\author{Kenji Watanabe}
\affiliation[NIMS]
{Research Center for Electronic and Optical Materials, National Institute for Materials Science, 1-1 Namiki, Tsukuba 305-0044, Japan}

\author{Takashi Taniguchi}
\affiliation[NIMS]
{Research Center for Materials Nanoarchitectonics, National Institute for Materials Science, 1-1 Namiki, Tsukuba 305-0044, Japan}

\author{Atindra Nath Pal}
\affiliation[S. N. Bose National Centre for Basic Sciences, Kolkata]
{S. N. Bose National Centre for Basic Sciences, Kolkata, West Bengal -700106,India }

\author{Vidya Kochat}
\email{vidya@matsc.iitkgp.ac.in}
\affiliation[Indian Institute of Technology, Kharagpur]
{Materials Science Centre, Indian Institute of Technology, Kharagpur, West Bengal - 721302,India }

\title[Skyrmionic Transport and First Order Phase Transitions in Twisted Bilayer Graphene Quantum Hall Ferromagnet]
  {Skyrmionic Transport and First Order Phase Transitions in Twisted Bilayer Graphene Quantum Hall Ferromagnet}

\begin{document}

%
%
%
%
%

\begin{abstract}
Large-angle twisted bilayer graphene (TBLG) realizes a multicomponent quantum Hall (QH) platform of spin, valley and layer pseudospins with strong Coulomb interaction-driven symmetry broken phases. Here, we investigate the low energy Landau-level spectrum of layer-decoupled TBLG and identify skyrmion-textured charged excitations and a field-induced insulating transition to an intervalley coherent state at zero-filling factor. Symmetric potential difference perpendicular to TBLG demonstrated layer coherent population of ground states with uniform energy barriers, while the charge imbalance in the layers at finite displacement field led to multidomain nucleation and a pronounced hysteresis in the exchange-dominated transport regime suggesting first order phase transitions between different QH ferromagnetic ground states.

\end{abstract}


\newpage
Multicomponent Quantum Hall (QH) physics arises from various degeneracies underlying the Hamiltonian of the material system resulting in additional pseudospin/isospin degrees of freedom such as the electron spin, valley index in multi-valley materials, the sub-band and layer index in 2D electron gas systems (2DEGS) with multiple quantum well structures or 2D bilayers \cite{1_spin,1_PRL1998,2_PRL1997,3_Science1998}. The spatial ordering of pseudospin/isospin degrees of freedom led to a new class of condensed matter systems termed as QH Ferromagnets \cite{10_science2000}. In modulation-doped semiconductor heterostructures and double quantum wells, as the interlayer separation ($d$) approaches the magnetic length ($l_B$), interlayer Coulomb coupling stabilizes excitonic, interlayer-coherent order and results in anomalous transport \cite{4_PRL2000,5_PRL1994,32_BEC_Bilayer,6_PRL2004,7_Science2000}. The strong Coulomb interaction in the lower LLs results in a large exchange energy favouring parallel spins, which dominates the Zeeman energy term and leads to energetically favourable skyrmion excitations observed in various 2DEGS and 2D bilayers \cite{8_PRB1993,9_science1995,4_book_Sarma1997PerspectivesIQ}. Similar QH ferromagnetic phases characterized by interaction-induced gaps and broken symmetries at integer filling factors have been observed in various graphene systems such as single/bi/tri/rhombohedral stacked multilayer graphene attributed to the spin-valley degeneracy in graphene \cite{12_Spin_valley_QHE,13_sym_break_BLG,14_QH_states,15_LLs_splitting,16_Broken_in_BLG,17_divergent_Resis,18_QHFM_trilayer,19_Broken_ABA_trilayer,20_tetralayer_graphene,21_Pentalayer_Graphene,5_2_QHF_BLG}.

The low energy band structure of monolayer graphene is described by two flavours of massless Dirac fermions at the inequivalent $K$ and $K'$ valleys of the hexagonal Brillouin zone giving rise to the anomalous QH sequence of $\sigma_{xy}=\pm\dfrac{4e^2}{h}\left(N+\dfrac{1}{2}\right)$ due to four-fold spin-valley degeneracy \cite{22_Graphene_discovery,23_QHE_graphene}. Thus the graphene LL is described by a single \textit{SU}(4) pseudospin and gives rise to ferromagnetic instabilities driven by exchange interactions, with order parameter corresponding to a finite polarization in spin and/or valley space. This results in finite charge excitation gaps at integer filling factors within the quartet LL manifested as QH pseudospin ferromagnetic states \cite{12_Spin_valley_QHE,24_QHFM_graphene,25_Skyrmion_SU4}. TBLG can be considered as a unique bilayer 2DEG with two mutually rotated single layers with interlayer Coulomb interactions and tunnelling in the extremity of a sub-nm tunnel barrier of 0.4 nm. At twist angles of ~1.05$^\circ$, called the magic angles, interlayer hybridization of Dirac cones produces flat bands, thereby leading to strong coupling of the layers \cite{40_Magic_angle_TBLG,41_Magic_angle_TBLG,42_magic_angle_TBLG}. At larger twist angles, the Dirac cones of each layer are significantly displaced in the momentum space  and this resultant momentum mismatch strongly suppresses interlayer tunnelling. In the large-angle limit, the layer quantum number is approximately conserved, so each Landau level becomes an eight-fold manifold (spin $\times$ valley $\times$ layer) that widens the set of competing pseudospin orders to include layer-polarized ferromagnets, interlayer-coherent excitonic condensates, valley-ordered textures, and mixed spin-valley states \cite{36_QHE_TBLG,007_robust_QHF_largeTBLG,33_30deg_TBLG,16_Broken_in_BLG}. Experimentally, large-angle devices show stable odd-integer interlayer-coherent states, a hierarchy of integer and fractional incompressible phases that respond to layer imbalance, and notably large gaps for interlayer-correlated phases \cite{34_OddInteger_QHS,007_robust_QHF_largeTBLG,008_Large_double_TBLG,11_1_Large_TBLG,11_2_Helical_Large_TBLG,11_3_Twist_dep_TBLG,11_4_militesla_LargeTBLG,70_Broken_symmetry_npj}. Here we present a systematic study of tilted-field and displacement-resolved QH transport in high-mobility single- and dual-gated large-angle TBLG focused on the interaction-driven symmetry breaking in the eightfold degenerate LLs of TBLG. Asymmetric screening produces uneven layer filling and multi-domain network of QH ferromagnetic ground states. Tilted field activation gap studies reveal the presence of charged spin/valley skyrmions in the low energy LL spectrum which transition from smooth skyrmion textures to metastable domian walls through first order phase transitions at finite displacement fields.

\begin{figure}[t]
  \includegraphics[width=\columnwidth]{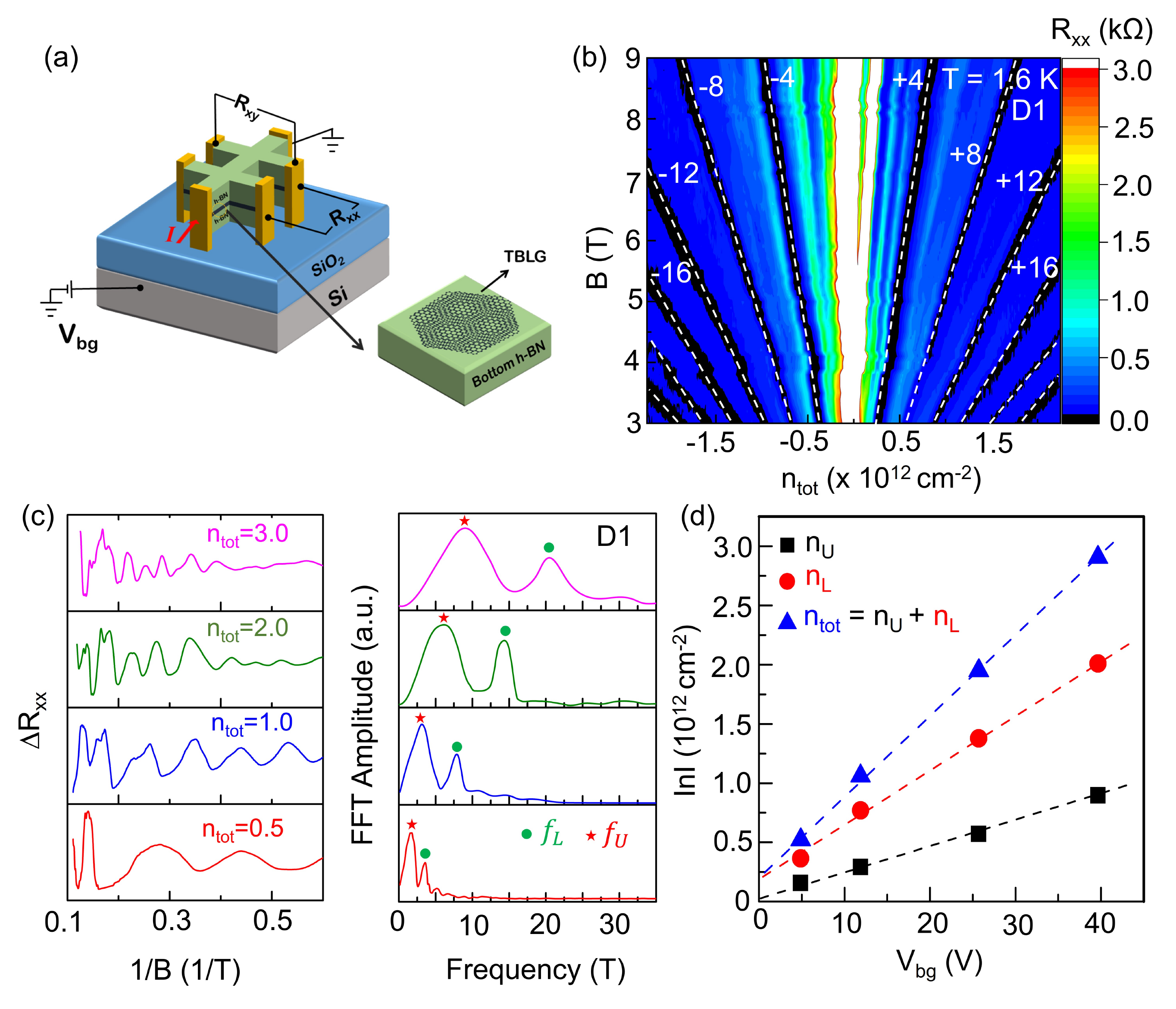}
\caption{\justifying(a) Schematic illustration of the bottom-gated h-BN encapsulated TBLG in Hall bar geometry.(b) Landau fan diagram showing \( R_{xx} \) as a function of \( B \) and  \( n_{\text{tot}} \) at 1.6 K. Different $\nu_{\text{tot}}$ values are shown in white dashed line.(c) \( \Delta R_{xx} \) vs. \( 1/B \) for different \( n_{\text{tot}} \) (\( \times 10^{12} \, \text{cm}^{-2} \)) values at 1.6 K (left) and corresponding FFT for each \( n_{\text{tot}} \) (right).(d) The upper layer (\( n_U \)), lower layer (\( n_L \)), and total (\( n_{\text{tot}} \)) carrier densities as function of gate voltage (\(V_{\mathrm{bg}}\)). The dashed lines are the theoretical fits of graphene double layer model to experimental  $|n|$ values estimated from FFT in (c).}
\phantomsubcaption\label{fig:1a} 
\phantomsubcaption\label{fig:1b}
\phantomsubcaption\label{fig:1c}
\phantomsubcaption\label{fig:1d}
\end{figure}

The large-angle TBLG devices in h-BN encapsulated Hall bar geometry were fabricated using the van der Waals pick-up and stack, followed by edge contact technique \cite{35_oneDime_Contact}. Randomly stacked TBLG heterostructures were characterized using Raman spectroscopy and the twist angles were estimated to be $\sim$5$^\circ$ and 8$^\circ$ for devices D1 and D2 respectively \cite{52_ACSNANO_VINEET,51_Raman_in_TBLG}. \ref{fig:1a} depicts the measurement configuration for bottom-gated h-BN/TBLG/h-BN device D1, while device D2 was fabricated as a dual-gated structure. Both the devices had mobilities of $\sim2,00,000$~cm$^2$/Vs and displayed well-resolved QH plateaus at magnetic fields as low as 1 T. The Landau fan diagram showing the evolution of $R_{xx}$ minima in magnetic field ($B$) along with back gate modulation of Fermi level, $E_F$ is shown in \ref{fig:1b} for D1. The integer QHE is observed with $R_{xx}$ minima at LL filling factors given by $\nu_{\text{tot}} = \pm4, \pm8, \pm12, \pm16,\dots$ indicating the presence of an eight-fold degenerate zero Landau level, with additional minima in the $N=0,1$ LLs arising from broken symmetry states. The low density of states in graphene leads to incomplete screening of the gate field by the bottom graphene layer, resulting in different charge carrier densities in the upper and lower graphene layers of D1. This results in a beating pattern in the Shubnikov-de Haas (SdH) oscillations measured in the $R_{xx}$ of D1 as illustrated in \ref{fig:1c}. The Fast Fourier Transform (FFT) of $\Delta R_{xx}$ ($R_{xx}$ oscillations centered around zero) vs. $1/B$ performed for different total carrier densities, $n_{tot}$ yields two prominent peaks corresponding to the two frequencies ($f_{U,L}$ for upper and lower layers) involved in the SdH oscillations as shown in \ref{fig:1c}. This translates to the carrier densities of upper and lower layers defined as $n_U$ and $n_L$ through the relation $n_{U,L}=gef_{U,L}/h$, where $g$ is the degeneracy factor of 4 arising from four-fold spin-valley degeneracy in each graphene layer, which is plotted in \ref{fig:1d} along with $n_{tot}$, where $n_{tot}=n_U+n_L$. These carrier density values can be fitted using a model of independently contacted graphene double layers separated by a thin dielectric \cite{55_SdH_Oscillations_in_CVD_TBLG,36_QHE_TBLG,58_Graphene_Capacitance_Cgg}. 
\begin{equation}
E_F(n_L)=\frac{e^2n_U}{C_{int}}+E_F(n_U)
\end{equation}
where $C_{int}$ is the interlayer capacitance of TBLG, whose value can be estimated from the fit to be  $5.7\pm1$ $\mu$F cm$^{-2}$ \cite{SM_PRL}.



\begin{figure}[t]
\includegraphics[width=\columnwidth]{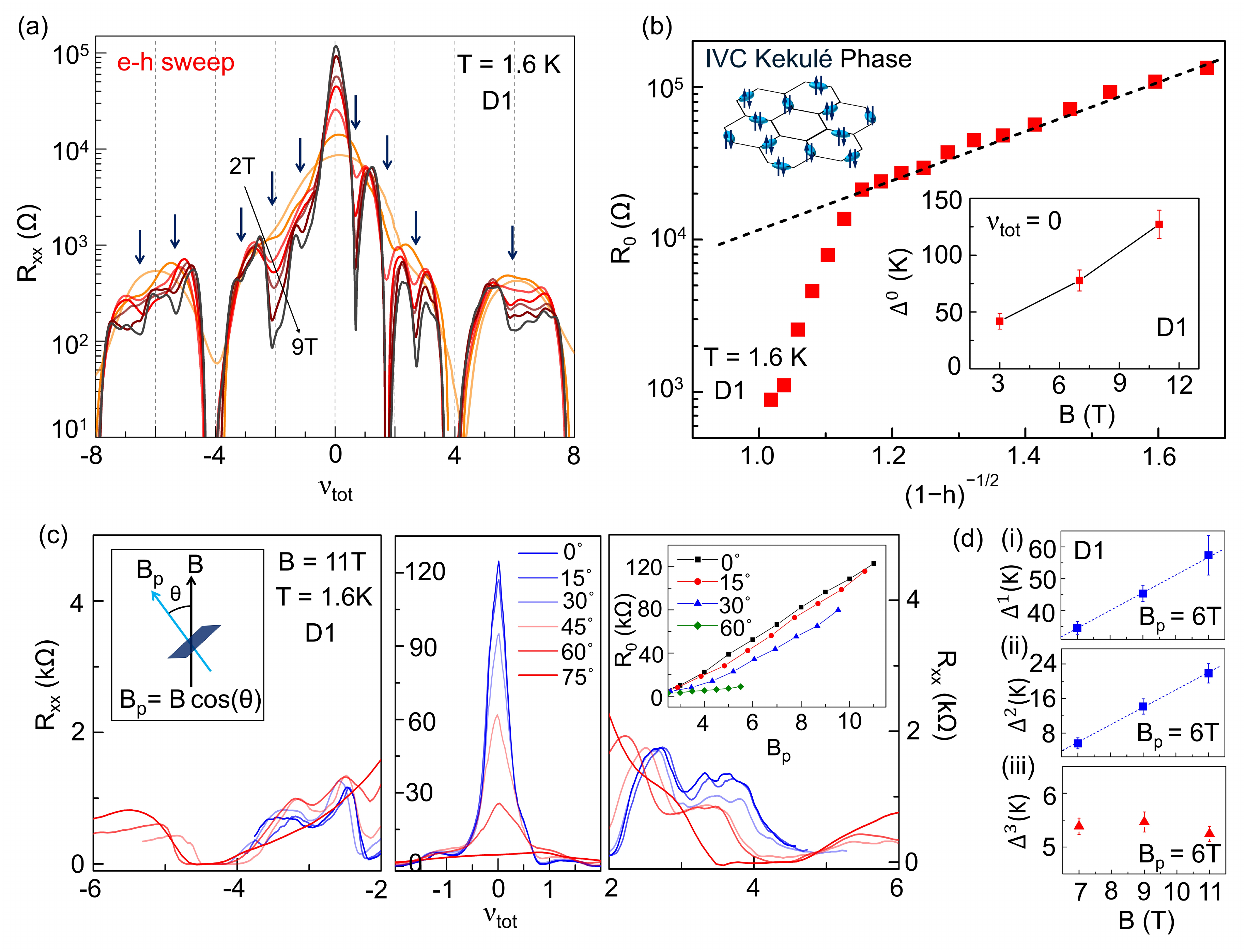}
\caption{\justifying
(a) Evolution of \(R_{xx}\) vs. $\nu_{\text{tot}}$ for e-h sweep at different B at 1.6 K with arrows indicating broken symmetry states. 
(b) \(R_0\) (i.e., \(R_{xx}\) at \(\nu_{\text{tot}} = 0\)) vs. \((1-h)^{-1/2}\) fitted using the KT equation (dashed line). 
Insets: Activation gap at \(\nu_{\text{tot}} = 0\), \(\Delta^0\) as a function of \(B\) and schematic of IVC Kekul\'e order phase. 
(c) \(R_{xx}\) vs. \(\nu_{\text{tot}}\) for various tilt angles \(\theta\) at \(B = 11~\text{T}\) as shown in left , center and right plots. Left inset: schematic of the tilted-field geometry. Righ inset: \(R_{0}\) vs. \(B_{p}\) at different \(\theta\).
(d)(i)-(iii) Activation gaps \(\Delta^\nu\) vs. \(B\) for \(\nu_{\text{tot}} = 1, 2, 3,\) at fixed \(B_{p} = 6~\text{T}\).  
The slope of the linear fit for each filling yields \(g_{\parallel}\) at the corresponding \(\nu_{\text{tot}}\).
}
\phantomsubcaption\label{fig:2a} 
\phantomsubcaption\label{fig:2b}
\phantomsubcaption\label{fig:2c}
\phantomsubcaption\label{fig:2d}
\end{figure}

In large-angle TBLG, the interlayer Coulomb interactions modulated by the carrier density in individual layers can compete with the magnetic field-tuned intralayer Coulomb interactions. This competition can result in new ground states and phase transitions in the QH ferromagnetic regime of TBLG.  \ref{fig:2a} shows the magnetic field evolution of the integer QH states as the Fermi level is tuned from electron to hole filling for D1 at 1.6~K. A magnetic field-induced QH insulator phase was observed at $\nu_{\text{tot}} = 0$ along with additional $R_{xx}$ minima within the eight-fold degenerate $N=0$ LL and at integer fillings in the higher LLs, signifying emergence of broken symmetry QH phases at high magnetic fields \cite{40_Phase_QHSs}. The QH insulator gap at $\nu_{\text{tot}} = 0$, $\Delta^0$ increases linearly with magnetic field $B$ as shown in \ref{fig:2b} inset, analogous to SLG \cite{44_Zero_Energystates}. In device D1 illustrated in \ref{fig:2b}, $R_0\sim\xi(h^2)$, with $R_0$ being the $R_{xx}$ at $\nu_{\text{tot}} = 0$ and $\xi$ the correlation length which has Kosterlitz-Thouless (KT) dependence given by \cite{43_KT_TBLG,45_KT_Theory}
\begin{equation}
\xi_{KT}\sim\exp(b/\sqrt{1-h})
\end{equation}
where $h=B/B_C$ with $B_C \sim 12 \,\text{T}$ \cite{SM_PRL}. In \ref{fig:2b}, the high-$B$ region is linear with slope, $b\sim0.79$, which agrees well with standard KT theory \cite{45_KT_Theory}. This also concurs with the recent scanning tunneling spectroscopic (STS) studies indicating magnetic field-tuned transition to an intervalley coherent state (IVC) with a Kekul\'e reconstruction, in which the LL associated with the $K$ and $K'$ valleys localize on B and A sublattices respectively \cite{69_Visualising_Kekule,39_QHBroken_Gr,40_Phase_QHSs,41_Canted_QHFM,42_Vanishing_Bulk_Heat}. In the high B limit, the ground state is an XY ordered phase formed from charge-neutral bound vortex-antivortex (V-AV) pairs, in which the Coulomb exchange favours fully valley-polarized state. Unbinding of V-AV pairs triggers KT transition to a disordered phase at low $B$ \cite{43_KT_TBLG}. Our experiment demonstrates that even in the limit of sub-nm interlayer spacing, each graphene layer in the TBLG preserves their individual pseudospin magnetic ground states at $\nu_{\mathrm{tot}}=0$. 

The tilted field studies of activation gaps at integer filling factors, $\Delta^\nu$, as shown in \ref{fig:2c} and \ref{fig:2d} helped to identify the signatures of spin textures of the ground states of the broken symmetry states in the $N=0$ LL. The spin responds to the total magnetic field, $B$, independent of the direction, unlike the orbital effects that depend only on the out-of-plane component, $B_p=B\cos\theta$ \cite{12_Spin_valley_QHE,014_Tilted_Field_Graphene,8_1_tiltedField,14_QH_states,15_LLs_splitting}. At a fixed $B$, increasing the tilt angle $\theta$ of the sample reduces the $B_p$ and weakens the gaps, which is shown in \ref{fig:2c} for $B=11$ T. The $R_0$ plotted as a function of $B_p$ estimated from a range of $B$ and $\theta$ values is shown in the inset of \ref{fig:2c}. From these traces we observe that $R_0$ decreases as the in-plane field grows (higher $\theta$) even though $B$ increases, indicating a spin-unpolarized insulating phase which is a further confirmation of the formation of IVC state at high $B$ \cite{12_Spin_valley_QHE}.  The calculated activation gaps extracted from the Arrhenius fit of \(\rho_{xx}\), $\Delta^{\nu}$, for the broken symmetry states at  $\nu_{\text{tot}} = 1$, $2$, $3$ as a function of $B$ is plotted in \ref{fig:2d} for a fixed $B_p$ of 6~T, by varying $\theta$. $\Delta^{1}$ and $\Delta^{2}$ exhibit a linear dependence on $B$, suggesting a spin-polarized ground state. Whereas, the negligible dependence of $\Delta^{3}$ on $B$ suggests spin-unpolarized, valley textured excitations. The $\Delta$ for $\nu_{\text{tot}} = 1,2$ can be fitted with the gap equation \cite{12_Spin_valley_QHE,012_skr_K-val,013_symbrk_sky_trns_TBLG,11_PRL1995,8_PRB1993} 
\begin{equation}
\Delta = \Delta_X(B_p, K) + (2K+1) \, g_0 \mu_B B - \Gamma 
\end{equation}  
where $\Delta_X(B_p,K)$ is the exchange energy and typically scales with the Coulomb energy $E_C\sim\displaystyle\frac{e^2}{4\pi\varepsilon\,\ell_B}$, $l_B$ $\propto1/\sqrt{B_p}$, $g_0$ is the bare Landé $g$-factor ($\approx 2$ in graphene), $\mu_B$ is the Bohr magneton and $\Gamma$ is the disorder broadening of the Landau levels. The second term is the Zeeman energy gap modified by the parameter $K$, where $K \ge 0$ represents the number of extra spins flipped in a skyrmion texture beyond the single spin flip of a conventional particle-hole excitation. For $K=0$, the excitation is a single spin flip and the slope gives $g_\parallel \approx g_0$, where $g_\parallel$ is defined as $g_\parallel\equiv\frac{1}{\mu_B}\partial_{B}\Delta$, whereas for $K>0$, multiple flipped spins lead to $g_\parallel > g_0$, a hallmark of exchange-dominated skyrmionic transport \cite{12_Spin_valley_QHE,013_symbrk_sky_trns_TBLG}. The fit to the measured values of $\Delta^{1,2}$ yielded $g_\parallel$ values of 7.92 and 5.96 for $\nu_{\text{tot}} = 1$ and 2 respectively, indicating that the lowest-energy excitations involve spin skyrmion textures. Our findings indicate that the low energy excitations in decoupled TBLG are of similar nature as observed in SLG,  where the lowest energy excitations are skyrmions in spin-valley space for $N\leq3$ where exchange energy dominates Zeeman anisotropy \cite{25_Skyrmion_SU4,51_magnon_Graphene}.

\begin{figure}[t]
\includegraphics[width=\columnwidth]{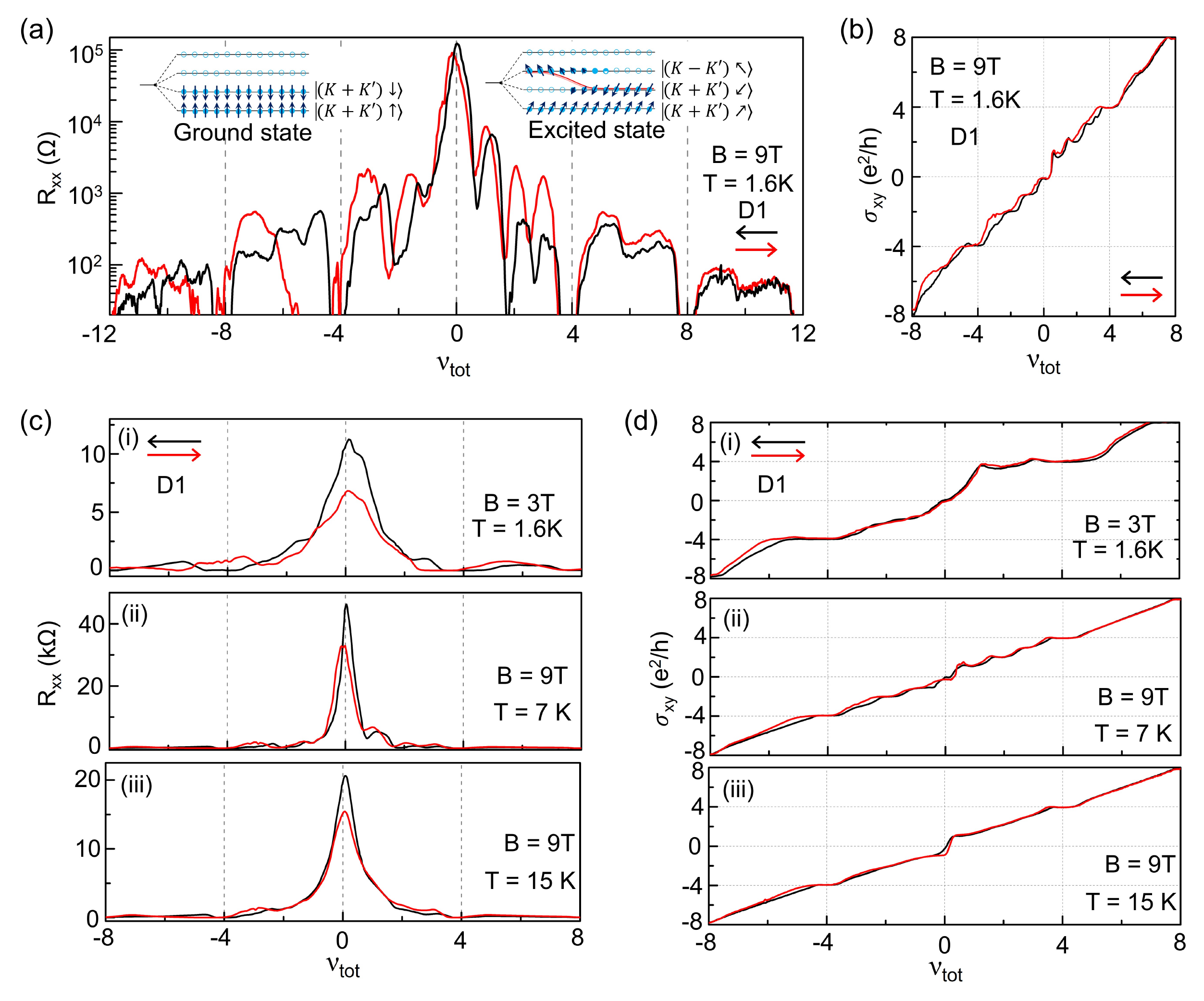}
\centering
\caption{\justifying
(a)-(b) Dual-sweep measurements of \( R_{xx} \) and $\sigma_{xy}$ vs. \( \nu_{\text{tot}} \) at \( B = 9 \,\text{T} \) and \( T = 1.6 \,\text{K} \). Black and red arrows indicate the direction of the voltage sweep. The insets in (a) show ground and excited states of the \(\nu_{\text{tot}} = 0\) IVC phase. (c) and (d) Dual-sweep measurements of \(R_{xx}\) and $\sigma_{xy}$ as a function of total filling factor \( \nu_{\text{tot}} \) for D1 under three conditions: (i) \(B=3\)\,T, \(T=1.6\)\,K (ii) \(B=9\)\,T, \(T=7\)\,K and (iii) \(B=9\)\,T, \(T=15\)\,K.
}
\phantomsubcaption\label{fig:3a} 
\phantomsubcaption\label{fig:3b}
\phantomsubcaption\label{fig:3c}
\phantomsubcaption\label{fig:3d}
\end{figure}

The evolution of integer QH states and the appearance of broken symmetry states was observed to depend remarkably on the direction of the gate voltage sweep direction as shown in \ref{fig:3a} and \ref{fig:3b}. The notable features here are the asymmetry in the emergence of broken symmetry states on the electron and hole sides as well as a considerable hysteresis which is developed in the lower LL regime where  $-8\leq\nu_{\text{tot}} \leq8$. We also note that the hysteresis develops only at low temperatures and high $B$, signified by the absence of hysteresis at low field of 3~T and at temperatures greater than 15~K as can be seen from the plots of $R_{xx}$ and hall conductance, $\sigma_{xy}$ as a function of total filling factor, $\nu_{\text{tot}}$ in \ref{fig:3c} and \ref{fig:3d}. The temperature scale where the hysteresis vanishes corresponds to the energy scale for pseudospin-pseudospin interaction energy which is of the order \(\sim 1.5\ \mathrm{meV}\) calculated from the hysteresis range.  This strongly suggests that the hysteresis is a consequence of the QH ferromagnetism developed in TBLG and is present in the QH regime of broken symmetry states where the transport is dominated by spin/valley skyrmion excitations. STS studies have shown direct observation of topological excitations of the IVC phase at $\nu_{\text{tot}}=0$ by visualizing valley skyrmions near charged defects, whose valley texture resembles canted antiferromagnetic skyrmion excitation of the Kekul\'e order \cite{39_QHBroken_Gr,69_Visualising_Kekule,49_Skyrmion_zoo}. In graphene LL at $\nu=\pm1$, corresponding to quarter filling, where Zeeman anisotropy is not dominant, all the electrons (or holes) spontaneously align their spin to reduce the Coulomb interaction. Instead of spin-flip transitions, this spin-polarization happens through formation of skyrmions carrying a charge of $\pm{e}$ due to their lower energy \cite{25_Skyrmion_SU4,12_Spin_valley_QHE}. In contrast, for $N\neq0$ with dominant Zeeman anisotropy, spin polarized ground states occur at half-filling and charged excitations with valley flip textures at quarter filling \cite{12_Spin_valley_QHE}.  The role of charge imbalance between the graphene layers and the skyrmion evolution as possible origins of QH hysteresis was further comprehensively explored in dual-gated structures.


\begin{figure}[t]
\includegraphics[width=\columnwidth]{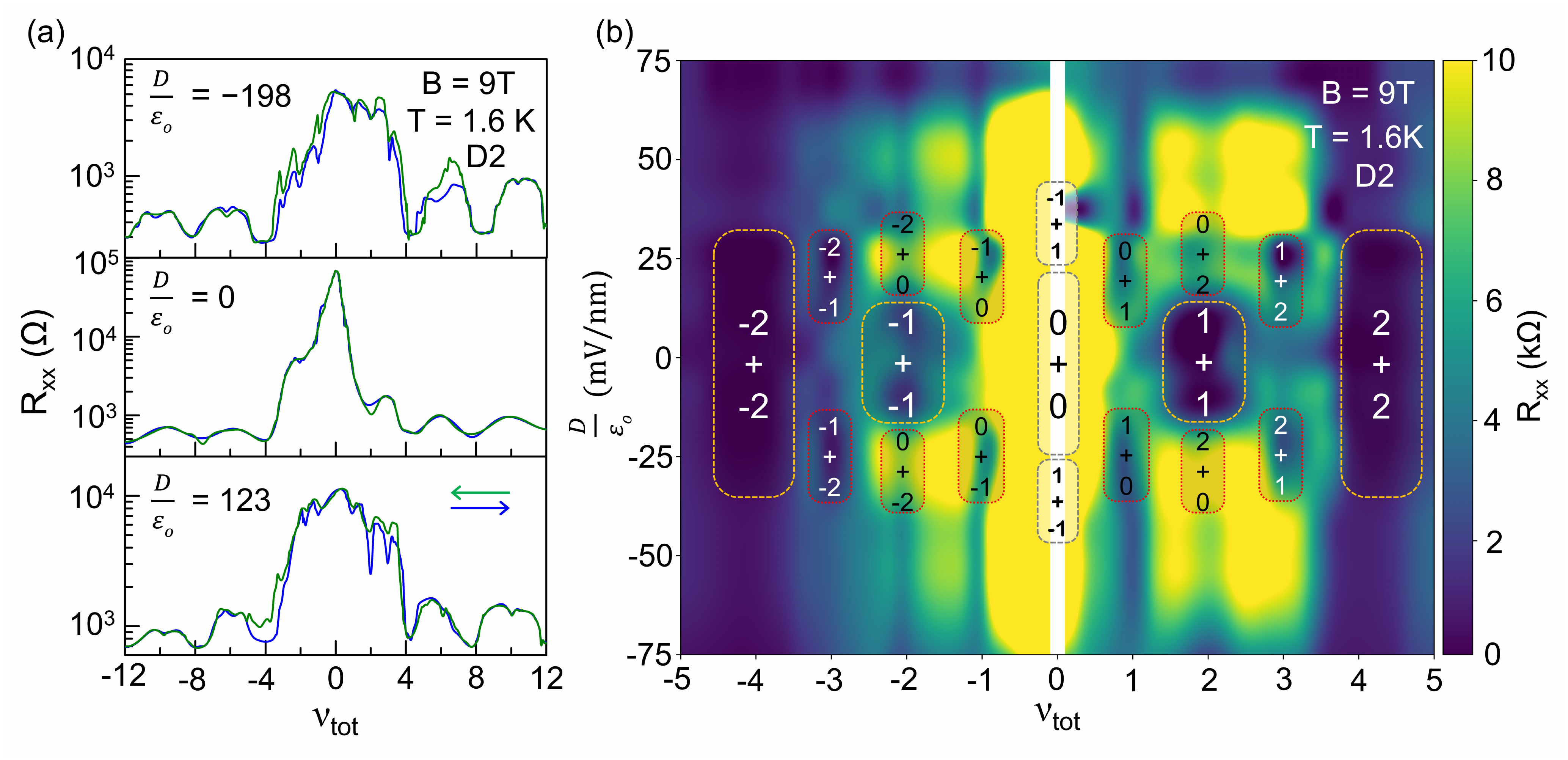}
\centering
\caption{\justifying
(a) Dual-sweep \(R_{xx}\) vs. \(\nu_{\text{tot}}\) at several \(D/\varepsilon_0\) (in unit of mV/nm) for D2 at \(B=9\)\,T and \(T=1.6\)\,K indicating broken symmetry states and hysteresis.(b) Contour plot of \(R_{xx}\) vs. \(D/\varepsilon_0\) and \(\nu_{\text{tot}}\) for D2 at \(B=9\)\,T and \(T=1.6\)\,K. The LL filling contributions from the upper and lower layers $(\nu_{U} + \nu_{L})$ are illustrated.
}
\phantomsubcaption\label{fig:4a} 
\phantomsubcaption\label{fig:4b}
\end{figure}

In dual-gated TBLG device, D2, the total charge carrier density and displacement field perpendicular to the layers can be independently controlled. Here the opposite gate sweep directions at 0~T showed no visible hysteresis at any $D$ \cite{SM_PRL}. At high values of $B$, QH ferromagnetic states were observed with marked differences between $D/\varepsilon_0=0$ and $|D/\varepsilon_0|\neq0$ as shown in \ref{fig:4a}, with $R_{xx}$ minima at $|\nu_{\text{tot}}|=1,3$ absent when $D/\varepsilon_0=0$, but emerges at finite $D$. At $D/\varepsilon_0=0$, when both the graphene layers are charge neutral ($\nu_{\text{tot}}=0$), we observe an insulating state that can be attributed to the IVC states in the individual graphene layers. Increasing the $\nu_{\text{tot}}$ at $D/\varepsilon_0=0$ results in filling of the LLs in the upper and lower graphene layers in a coherent manner, resulting in $R_{xx}$ minima at $\nu_{\text{tot}}=2,4$ corresponding to quarter-filled and fully-filled LLs respectively of individual layers.  Applying a perpendicular $D$ explicitly breaks layer degeneracy and produces an easy-axis anisotropy that energetically favors one layer, where the intra-layer exchange readily polarizes the spin-valley flavours in the favoured layer and odd fillings at $\nu_{\text{tot}}=1,3$ appear \cite{007_robust_QHF_largeTBLG,34_OddInteger_QHS,11_2_Helical_Large_TBLG}. The full spectrum of broken symmetry states observed in the  filling of lowest LL in TBLG system is shown in \ref{fig:4b}. Another interesting observation here is that the hysteresis is completely absent when $D/\varepsilon_0=0$, but increases with $D$ and can be observed till $\nu_{\text{tot}}=12$ at the highest accessible $D$ in D2 device. The pseudospin polarization in the spin-valley space of the graphene layers leads to Ising ferromagnetism, with inevitable nucleation of multiple domains in the presence of disorder or at finite temperature \cite{6_First_order_Phase,12_nanoletters2017}.  When both layers are filled coherently as for the case of $D/\varepsilon_0=0$, the ferromagnetic ground states as well as the lowest energy excitations in both layers are analogous forming a uniform potential landscape. This favours continuous rotation of the pseudospin polarization without creating sharp domain boundaries and with reversible reconfiguration with opposite gate sweep directions. A finite displacement field results in charge imbalance between the layers that alters the energy landscape in the layers differently through screening of the residual disorder in individual layers. This results in the formation of pseudospin ordered domains and disorder nucleated domains of low energy excitations of skyrmions, separated by domain walls, where the pseudospin index (spin, valley) is different in the sublayers \cite{010_Hysteretic_TransportGap_BLG}. The hysteresis observed at finite $D$ indicates first-order phase transitions between the various magnetic ground states as a particular broken symmetry state is approached \cite{6_First_order_Phase,7_Science2000,13_PRX2023,14_PRL1998,10_science2000}. It specifically depends on the initial pseudospin index configuration of the sublayers which have energetically different anisotropy barriers during opposite gate sweep directions, analogous to a memory effect based on the history of gate sweep direction. The hysteretic behaviour is observed only in exchange-dominated transport regime with skyrmionic low energy excitations whose size scales down with increasing disorder that weakens long-range spin/pseudospin interactions. Further evidence is provided by the absence of hysteresis in a device, D3 with relatively larger disorder, at field of 9~T and 1.6~K \cite{SM_PRL}. Here the QH ferromagnetic states were not visible due to larger disorder strength which rapidly shrinks the domain sizes under similar experimental conditions.

In conclusion, we observe the interplay of charge screening effects, broken symmetry states and multidomain physics in the QH ferromagnetic regime of decoupled TBLG. The exchange interaction energy scale dominates the low energy transport, resulting in ferromagnetic QH ground states with lowest energy excitations of skyrmion texture in spin-valley pseudospin space as inferred from tilted field activation gaps. Investigation of dual-gated TBLG devices demonstrated layer coherent population of ground states with uniform energy barriers at zero displacement fields, while the charge imbalance in the layers at finite displacement field led to multidomain nucleation and hysteresis between different QH ferromagnetic ground states driven by first order phase transitions. Our work paves way for understanding symmetry breaking phases in distinctly stacked multilayer graphene and Moir\'e systems.

\section{Acknowledgement}

V.P., P.G., S.P., A.M.B and V.K. acknowledge funding support from the DST-Nanomission programme of the Department of Science and Technology, Government of India (DST/NM/TUE/QM-1/2019) and the STEP facility, IIT Kharagpur. R.P. and A.N.P. acknowledge the Thematic Unit of Excellence on Nanodevice Technology (grant no. SR/NM/NS-09/2011) and the Technical Research Centre (TRC) Instrument facilities of S. N. Bose National Centre for Basic Sciences, established under the TRC project of Department of Science and Technology (DST), Govt. of India. A.N.P. acknowledges DST Nano Mission: DST/NM/TUE/QM-10/2019. K.W. and T.T. acknowledge support from the JSPS KAKENHI (Grant Numbers 21H05233 and 23H02052) , the CREST (JPMJCR24A5), JST and World Premier International Research Center Initiative (WPI), MEXT, Japan.

\bibliography{references}

\clearpage
\includepdf[pages=-]{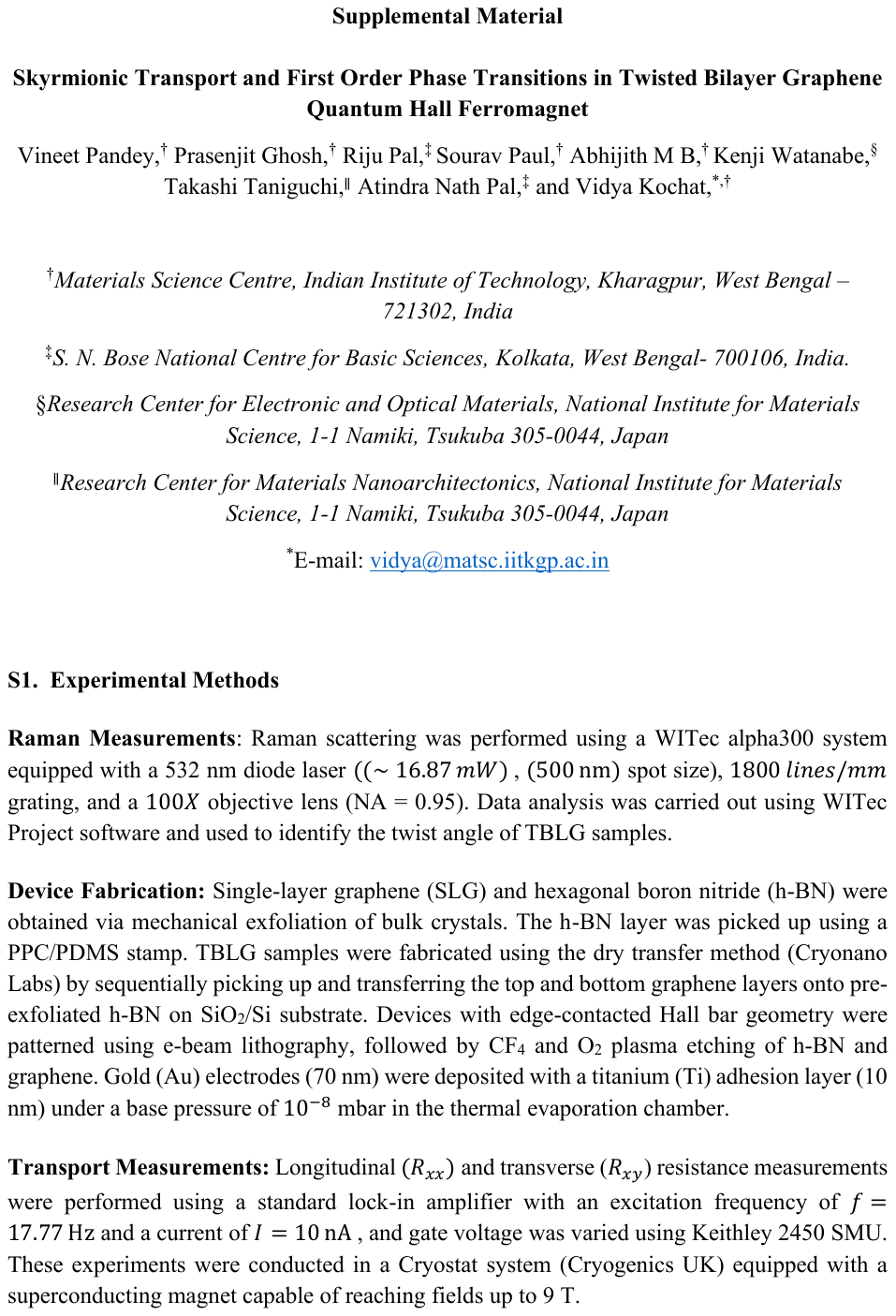}

\end{document}